\documentclass{pasj00}

\usepackage{color}
\usepackage{pst-grad}
\usepackage{psfrag}
\usepackage{url}

\begin{document}
%\Received{$\langle$reception date$\rangle$}
%\Accepted{$\langle$acception date$\rangle$}
%\Published{$\langle$publication date$\rangle$}
\SetRunningHead{Astronomical Society of Japan}{}

\newcommand{\vdag}{(v)^\dagger}
\newcommand{\Gray}{$\gamma$-ray~}
\newcommand{\Grays}{$\gamma$-rays~}
\newcommand{\mf}{100~$\mu$G~}
\newcommand{\hii}{H{\sc ii} region}

%%%%%%%%%%%%%%%%%%%%%%%%%%%%%%%%%%%%%%%%
\title{Molecular Clouds as Cosmic-ray Barometers}

%%% begin:list of authors
% Do NOT capitalize all letters in "textsc".
\author{Sabrina \textsc{Casanova} \altaffilmark{1},
          \email{sabrina.casanova@mpi-hd.mpg.de}
        Felix A.  \textsc {Aharonian}\altaffilmark{1,2},
        Yasuo  \textsc {Fukui}\altaffilmark{3},
        Stefano \textsc {Gabici}\altaffilmark{2},
        David I. \textsc{Jones} \altaffilmark{1},
        Akiko \textsc {Kawamura}\altaffilmark{3},
        Toshikazu \textsc{Onishi}\altaffilmark{3},
        Gavin \textsc {Rowell}\altaffilmark{4},
        Kazafumi \textsc {Torii} \altaffilmark{3},
        Hiroaki \textsc {Yamamoto}\altaffilmark{3}
}

\altaffiltext{1}{Max Planck f\"ur Kernphysik, Saupfercheckweg 1, 69117, Heidelberg}
\altaffiltext{2}{Dublin Institute for Advance Physics, 31 Fitzwilliam Place, Dublin 2, Ireland}
\altaffiltext{3}{Nagoya University, Furo-cho, Chikusa-ku, Nagoya City, Aichi Prefecture, Japan}
\altaffiltext{4}{School of Chemistry and Physics, University of Adelaide, Adelaide 5005, Australia}

%%% end:list of authors
%% `\KeyWords{}' always has to be placed before `\maketitle'.

 %Do NOT move this preamble from here!

\maketitle

\begin{abstract}

The advent of high sensitivity, 
high resolution \Gray detectors, together with a knowledge of the 
distribution of the atomic hydrogen 
and especially of the molecular hydrogen in the Galaxy on sub-degree 
scales creates a unique opportunity 
to explore the flux of cosmic rays in the Galaxy. We here present 
the new data on the distribution of the molecular hydrogen 
from a large region of the inner Galaxy obtained by the NANTEN Collaboration. 
We then introduce a methodology which aims to provide a test bed 
for current and future \Gray observatories to 
explore the cosmic ray flux at various positions in our Galaxy. 
In particular, for a 
distribution of molecular clouds, as provided by the NANTEN survey, 
and local cosmic ray density as measured at the Earth, 
we estimate the expected GeV to TeV \Gray signal, which can then be 
compared with observations and use to test the cosmic ray flux. 

\end{abstract}

\section{Introduction}\label{sec:intro}

The question of the distribution of the cosmic rays (CRs) 
and of the CR sources in the Galaxy was addressed extensively 
by many authors (see for instance \cite{Ginzburg}). 
Secondary CR data tell us that 
CR protons and nuclei, that are accelerated by Galactic sources, have a confinement time in the Galaxy of about 
$10^7$~years at GeV energies, which decreases with increasing energy. 
During this time the particles accelerated by individual sources mix together, lose memory of their origin, 
and contribute to the bulk of Galactic CRs known as the cosmic ray {\it sea} or CR background. 

{ Direct measurements provide the CR spectrum and flux only 
in the vicinity of the Solar System and the level of the CR spectrum and 
flux in other regions of the Galaxy is unknown. In order to investigate 
the CR level in distant regions of the Galaxy 
and the distribution of Galactic CR sources one can start assuming that the 
CR flux and spectrum measured locally is representative
of the typical CR flux and spectrum present throughout the Galaxy,
except in the proximity of isolated CR sources. The implications of such 
an assumption are} firstly, 
that there
exists a definable CR {\it sea} of homogenised and isotropised CRs, 
far from any of the kpc-scale uniformly and continuously distributed CR sources, 
which is the typical CR flux and spectrum in the Galaxy; and secondly, that the Earth 
sits in a typical region far from any source where this {\it sea} 
can be measured directly. Although the ${10}^7$ years diffusion time of 
CRs in the Galaxy seems to imply the existence of the {\it sea}, in principle 
there could indeed be a {\it sea} of CRs in the Galaxy, 
but the Earth could be located close to a so far undetected CR source, 
so that the locally measured flux is therefore substantially higher 
than the {\it sea} and the spectrum substantially harder, and the 
local values are therefore not representative of the typical case. 
Alternatively, if the CR sources are 
distributed quite densely, on sub-kpc scales in the Galaxy, there may not really be a {\it sea} 
CR spectrum at all. The local values of CR flux and spectrum would then vary 
strongly depending mostly 
on the proximity to the CR sources in any particular region. 
In this case the flux value at the Earth could be lower or higher, and the spectrum
softer or harder, than any given distant region.
Finally, it could be that 
the CR sources are so densely distributed in the Galaxy, that there might exist 
strong gradients in the CR spectrum, 
depending mostly on the proximity to the CR sources in any particular region, 
and the Earth might by chance be located unusually far from any source and 
thus the locally measured
flux and spectrum are lower and softer than typical values found elsewhere.

Whilst diffusing through the Galaxy, CRs interact with ambient matter (both molecular and atomic) and produce \Grays 
through inelastic collisions. This emission has long been recognised by many authors as a 
unique probe of the parent CR distribution because the resultant 
\Gray emissivity is a function only of the matter density and the CR spectrum 
\citep{Ginzburg,Puget,Casse,BloemenB,BloemenC,Lebrun,Bertsch,Hunter,Strong}. 
For instance \cite{BloemenC} investigated the galacto-centric distribution of cosmic rays. 
They defined four intervals 
in the galacto-centric radius, for which they calculated the column densities 
of atomic and molecular hydrogen. By comparing the $\gamma$-ray intensities 
with the predicted emissivities, they derived an exponential distribution 
for the cosmic rays in the Galaxy. In the paper by \cite{Bertsch} 
the CR flux used for calculating the $\gamma$-ray flux is varied position to 
position by multiplying by a "cosmic ray enhancement factor" 
$c(r,l,b)$, which is derived 
by convolving the matter distribution with a Gaussian 
distribution whose width is an adjustable parameter $r_0$. The cosmic ray enhancement factor 
$c(r,l,b)$ takes a value ranging from 0.1 to 1.6. The CR 
flux in the Galaxy varies significantly to that measured 
in the solar neighbourhood. 
The most comprehensive numerical model, which includes particle production
and propagation in the Galaxy, is GALPROP \citep{Strong} {\footnote{GALPROP is available 
from: \url{http://galprop.stanford.edu/}}}. GALPROP is designed to
perform cosmic-ray propagation calculations for nuclei, antiprotons, electrons and positrons, and
computes diffuse $\gamma$-rays and synchrotron emission. In GALPROP the 
CR propagation equation is solved numerically on a spatial grid, either in 2D with
cylindrical symmetry in the Galaxy or in full 3D. The numerical solution proceeds in
time until a steady-state is reached. Normalisation of protons, 
helium and electrons to experimental data 
is provided (all other isotopes are determined by the source composition and 
propagation). $\gamma$-rays and synchrotron are computed using interstellar gas data 
(for pion-decay and bremsstrahlung) and the interstellar radiation fields (ISRF) model (for inverse Compton). 
Spectra of all species on the chosen grid and the $\gamma$-ray and synchrotron sky-maps 
are output in a standard astronomical format for comparison with data. The approach adopted by \cite{Strong} 
to include effects of CR flux variations in the Galaxy is to calculate the 
propagation effect from CR sources, which results in variation of CR flux in our Galaxy \citep{moskalenko}. 
The simulated $\gamma$-ray flux is then the
background, which must be subtracted to recognise the sites of concentrated excess $\gamma$-ray
emission in a detector data, such as in Fermi data. 

Interestingly, based on EGRET data \citep{Hunter} 
the \Gray emission from the inner Galaxy 
(${300}\,^{\circ}<l<60\,^{\circ}$ and ${-10}\,^{\circ}<b<10\,^{\circ}$, as defined in \cite{Hunter}) 
at energies greater than about 1~GeV exceeds, by some 60$\%$,  
the intensity predicted by model calculations { \cite{Bertsch,Hunter}}. 
Recent observations, performed by the LAT instrument on board the 
Fermi satellite, show that the spectra of the Galactic 
diffuse emission at MeV-GeV energies, at least at intermediate latitudes \citep{Ackerman}, 
can be explained by cosmic-ray propagation models based on local observations of 
cosmic-ray electron and nuclei spectra. On the other hand, at TeV energies the spectral features of the \Gray emission detected by HESS 
from the Galactic centre (GC) region \citep{Aharonian:nature}, and the emission measured by Milagro from the Cygnus Region 
\citep{Abdo:2007,Abdo:2008}, suggest that the CR flux might significantly vary in the different locations of the Galaxy.  

In this paper we seek to compare the CR flux and spectrum observed at the
Earth to the CR flux and spectrum in specific distant regions, derived from observations 
of \Grays emitted by CRs propagating through molecular clouds. We first 
present not publicly available CO observations obtained with the NANTEN telescope for 
a significant and interesting part of the galactic plane. We then demonstrate that the NANTEN data 
in combination with the Fermi data allow meaningful studies of deviations of the cosmic ray flux 
from the average one. The level of the {\it sea} of lower energy CRs ($<$ few GeV) has been 
traced by the ionisation rate inferred from different 
molecular species seen in molecular clouds \citep{Indriolo2007,Indriolo2009}. 
The level of the CR {\it sea} above 1 GeV will then be inferred 
by combining the \Gray emission from molecular clouds 
with the data from the Leiden/Argentine/Bonn (LAB) survey of the 
Galactic atomic hydrogen \citep{Kalberla} 
and from the NANTEN survey of Galactic molecular hydrogen \citep{Fukui1, fukui4, Fukui2, Fukui3}. 
Studying the $\gamma$-ray emission from single molecular clouds 
the differential, local distribution of the cosmic ray flux can be inferred. 
The detection of even a single under-luminous 
cloud with respect to the predictions based on the local CR density 
would cast severe doubt on the assumption that the level of the observed 
CR flux is representative of the average CR flux and spectrum 
in the Galaxy. 

In Section \ref{sec:interactions} we will present new data on the distribution of the molecular hydrogen 
from the region of the inner Galaxy, which spans Galactic longitude $340^\circ<l<350^\circ$ and 
Galactic latitude $-5^\circ<b<5^\circ$, obtained by the NANTEN Collaboration. 
We will also briefly review the results 
of the LAB survey of the Galactic atomic hydrogen. A description of our approach will be given in Section 
\ref{sec:testing}. The method will be here applied to the region of the Galactic Plane described above 
to provide a 'proof-of-concept'. A more 
comprehensive study on a larger scale will be discussed in a future paper. 
Some predictions concerning the \Gray and cosmic ray fluxes from molecular clouds including their uncertainties 
are given in Section \ref{sec:results}, where we will also discuss the observational prospects for present and 
future observatories. Our conclusions are given in Section \ref{sec:conclusions}.      

\section{The gas distribution}\label{sec:interactions}

\subsection{The NANTEN survey of molecular hydrogen}\label{sec:NANTEN}
The NANTEN instrument is a 4~m millimetre/sub-millimetre telescope which has surveyed the southern sky 
using the  $^{12}$CO (J=1--0) emission line at 115.271~GHz ($\lambda=2.6$~mm). The angular resolution at 
this frequency is 4$'$, with a mass sensitivity of about 100~M$_\odot$ at the Galactic centre (GC), assuming a 
distance of 8.5~kpc for the GC. The NANTEN data set has 1~km/s resolution in velocity \citep{fukui4,Fukui1,Fukui2,Fukui3}.  
The conversion factor, $X$, which is used to translate the intensity of CO molecular line emission into H$_2$ column 
density \citep{Nakanishi1,Nakanishi2}, is assumed to be
 \begin{equation}\label{eq:X}
 X= 1.4 \times {10}^{20}  \, e^{(R/11 \, {\rm kpc})}{\rm {~[cm}^{-2} {K}^{-1} {km}^{-1} s]},
 \end{equation} 
where $R$ is distance to the GC. This conversion factor, which is based on the metalicity distribution in the Galaxy, 
is a function of the distance to the GC. 
The uncertainty in the determination of the distance to the GC and in the determination of the calibration itself 
leads to an uncertainty in $X$ of a factor of 50$\%$ \citep{arimoto}. 

The final data product for the region, which spans Galactic longitude $340^\circ<l<350^\circ$ and Galactic latitude 
$-5^\circ<b<5^\circ$, consists of a $150\times150\times600$ element three-dimensional image of the molecular hydrogen 
distribution as a function of Galactic longitude, latitude and velocity. Assuming a flat rotation curve model of the Galaxy 
with uniform velocity equal to $220 \, {\rm km/s}$, the data-cubes in 
longitude, latitude and velocity are 
transformed to data-cubes in longitude, latitude and heliocentric distance 
\citep{Nakanishi1,Nakanishi2}. 

\begin{figure}
\includegraphics[width=0.48\textwidth]{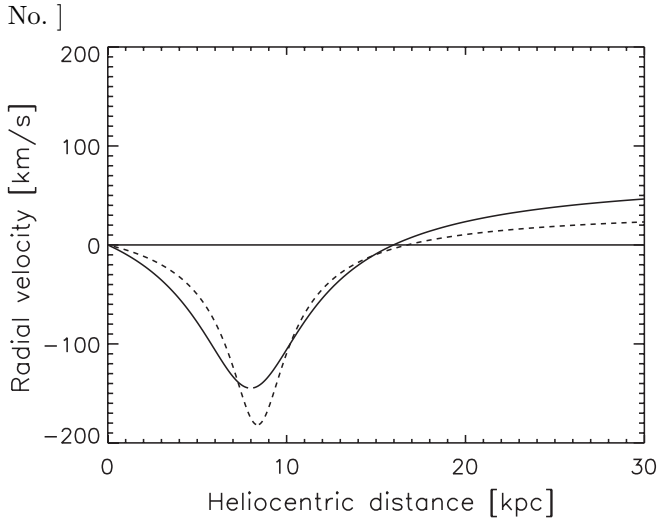}
\caption{ The radial velocity vs heliocentric distance estimated from flat rotation of the galactic disk is shown. 
The solid line is at (l,b)~(340,0) and the dashed line is at (l,b)~(350,0). \label{fig1} } 
\end{figure}

Figure \ref{fig1} shows a plot of the radial velocity versus heliocentric distance used 
to convert the velocity information into distance. There are two distance 
solutions for each radial velocity in the inner Galaxy. In order to resolve this 
near-far distance ambiguity, the method by \citep{Nakanishi1,Nakanishi2} is used. 
The principle uncertainty in the determination of the distances in the 
NANTEN data-set comes from errors in the accuracy of the radial velocity estimates. 
The combination of turbulence and peculiar 
proper motions of the gas causes spectral line broadening, which 
introduces uncertainties in the radial velocity, that can be as large 
as 15~km/s (a giant molecular cloud typically shows velocity width of 10 km/s, 
and the cloud-cloud motion is 3-3.9 km/s \citep{alvarez,clemens}). 
This corresponds to an uncertainty of less than 2 kpc in the distance. 
The dependence of the uncertainties in the radial velocity on the particular location 
is small for our dataset, since the longitude range used here 
includes disk gas, and excludes the Galactic centre and molecular 
ring. For disk gas, it is, in fact, unlikely to have significant systematic 
variation in either cloud-cloud velocity dispersion or typical 
line widths with location. Correspondingly, the distance error arising 
from molecular gas velocity dispersion is likely to be fairly constant 
throughout our data sample.

\subsubsection{Gas morphology}\label{sec:GM}

Figure \ref{fig2}~({\it Left}) and \ref{fig2}~({\it Right}) 
show NANTEN observations of the $^{12}$CO (J=1--0) emission line from a large area of the Galactic plane. 
This segment spans Galactic longitude $340^\circ<l<350^\circ$ and Galactic latitude $-5^\circ<b<5^\circ$. 
Figure \ref{fig2}~({\it Left}) shows the integrated emissivity in units of cm$^{-3}$~km~s$^{-1}$ 
from this region. Inspection of Figure \ref{fig2}~({\it Left}) shows that the molecular material is predominantly confined to 
$\lesssim2^\circ$ of the plane.  The longitude span places this region in between the Galactic centre and the 
Norma arm \citep{Duncan1995} so that there should be no contamination from depth effects from looking down an arm. 
This is reflected in the distribution of heliocentric distance in the region in Figure \ref{fig2}~({\it Right}), 
which shows the distance as a function of position. Figure \ref{fig2}~({\it Right}) shows, in fact, a moment 1 map of the 
region, which is $\sum[I \, d]/\sum[I]$, where $I$ is the intensity and $d$ is the distance of the gas and has units of 
kilo-parsec. There are a few regions of emission which are very far away (corresponding to the dark pixels). 
However, these tend to be isolated, small segments on the edge of the plane, and hence they should contribute a 
negligible amount to the \Gray emissivity.  Specifically, Figure \ref{fig2}~({\it Right}) shows that the bulk of 
the observed emission originates at distances of $\lesssim2$~kpc, whilst there is some 
emission, mainly confined to the central regions of the Galactic plane, which are at greater distances. 
 
\subsection{The Leiden/Argentine/Bonn Galactic HI Survey}\label{sec:LAB}

We use data from the Leiden/Argentine/Bonn (LAB) Galactic HI Survey 
\footnote{available from:\\\url{http://www.astro.uni-bonn.de/~webaiub/english/tools_labsurvey.php}} 
\citep{Kalberla} for the distribution of the atomic hydrogen. 
The LAB observations were centred on the  
1420~MHz ($\lambda21$~cm) line with a bandwidth of 5~MHz. The 
velocity axis spans $-450$~km~s$^{-1}$ to $400$~km~s$^{-1}$ giving a 
final velocity resolution of $1.3$~km~s$^{-1}$.  The data combines 
observations from three telescopes at a common resolution of $0.6^\circ$ 
with a final sensitivity of $\sim0.1$~K.  We used the same procedure 
described in \ref{sec:NANTEN} to transform the velocity axis into distance.

\section{Testing the level of the CR background}\label{sec:testing}

The emissivity of molecular clouds located far away from CR sources, 
so called {\it passive} molecular clouds, i.e. clouds which are 
illuminated by the supposedly existing CR background, can be used to probe the level
of the CR {\it sea} \citep{Issa}. Given that  the \Gray-emission from 
the molecular cloud depends only upon the total mass of the cloud, 
M, and its distance from the Earth, $d$, the CR flux, $\Phi_{CR}$, in the cloud is uniquely 
determined as
\begin{equation}
 \Phi_{CR} \propto \frac{{F_{\gamma}} \, d^2} {M} 
\label{eqn:fluxCloud}
\end{equation} 
where ${F_{\gamma}}$ is the integral \Gray flux from the cloud and $M=n\,V$, with $n$ the gas number density and $V$ the volume of the cloud. 
Under the assumption that the CR background is equal to the locally observed CR flux, the calculated $\gamma$-ray flux from the cloud 
can be compared to the observed $\gamma$-ray flux in order
to probe the CR spectrum in distant regions of the Galaxy. The detection of under-luminous clouds with 
the respect to predictions based on the CR flux at Earth would 
suggest that the local CR density is enhanced 
with respect to the Galactic average density. This would cast doubts 
on the assumption that the local CRs are produced only by distant sources, and 
that the CR flux and spectrum measured locally is representative
of the typical CR flux and spectrum present throughout the
Galaxy \citep{Aharonian:2000}.

\subsection{Diffusion of CRs into molecular clouds}

\begin{figure*}
\begin{center}
\includegraphics[width=0.48\textwidth]{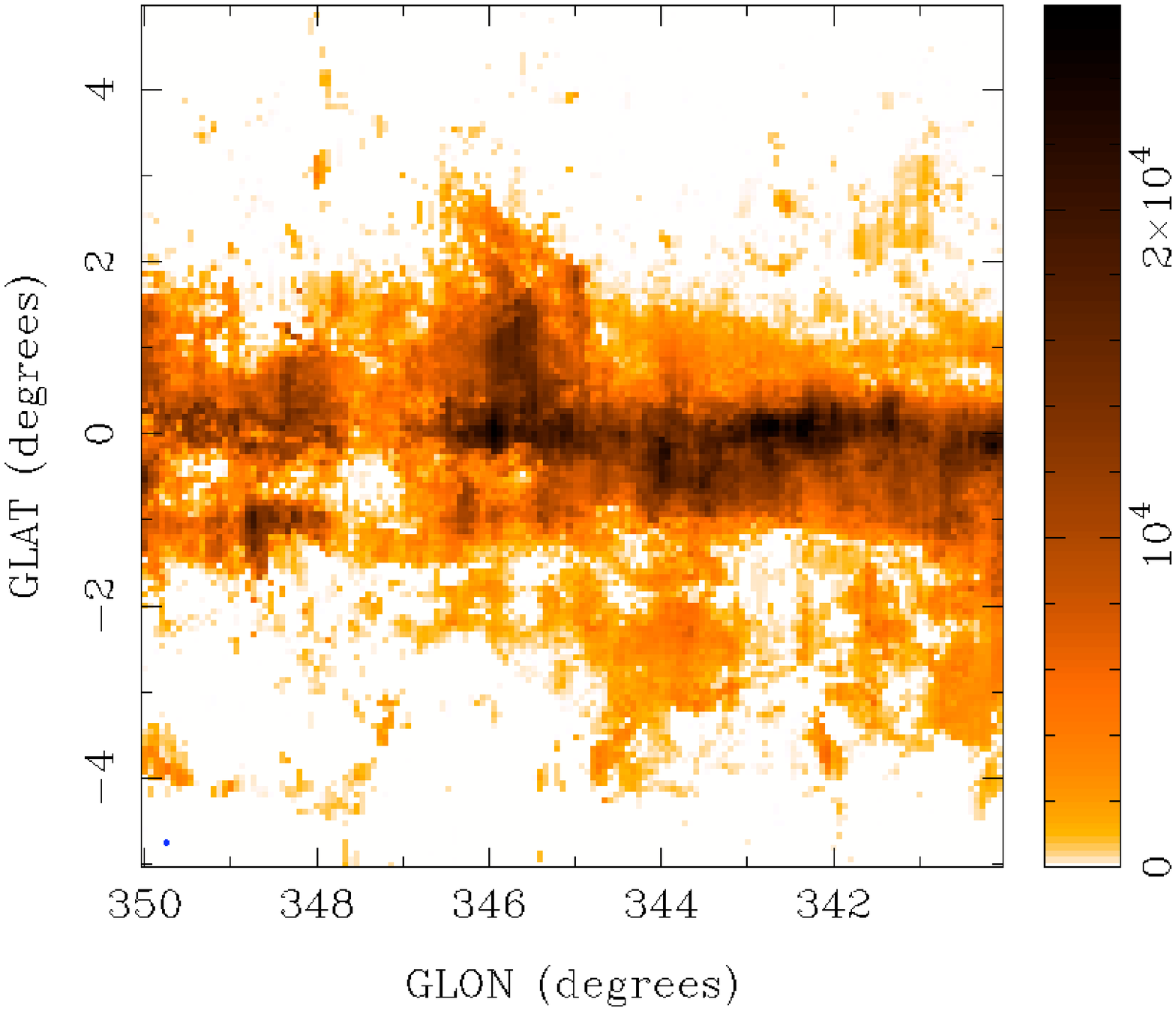}
\includegraphics[width=0.48\textwidth]{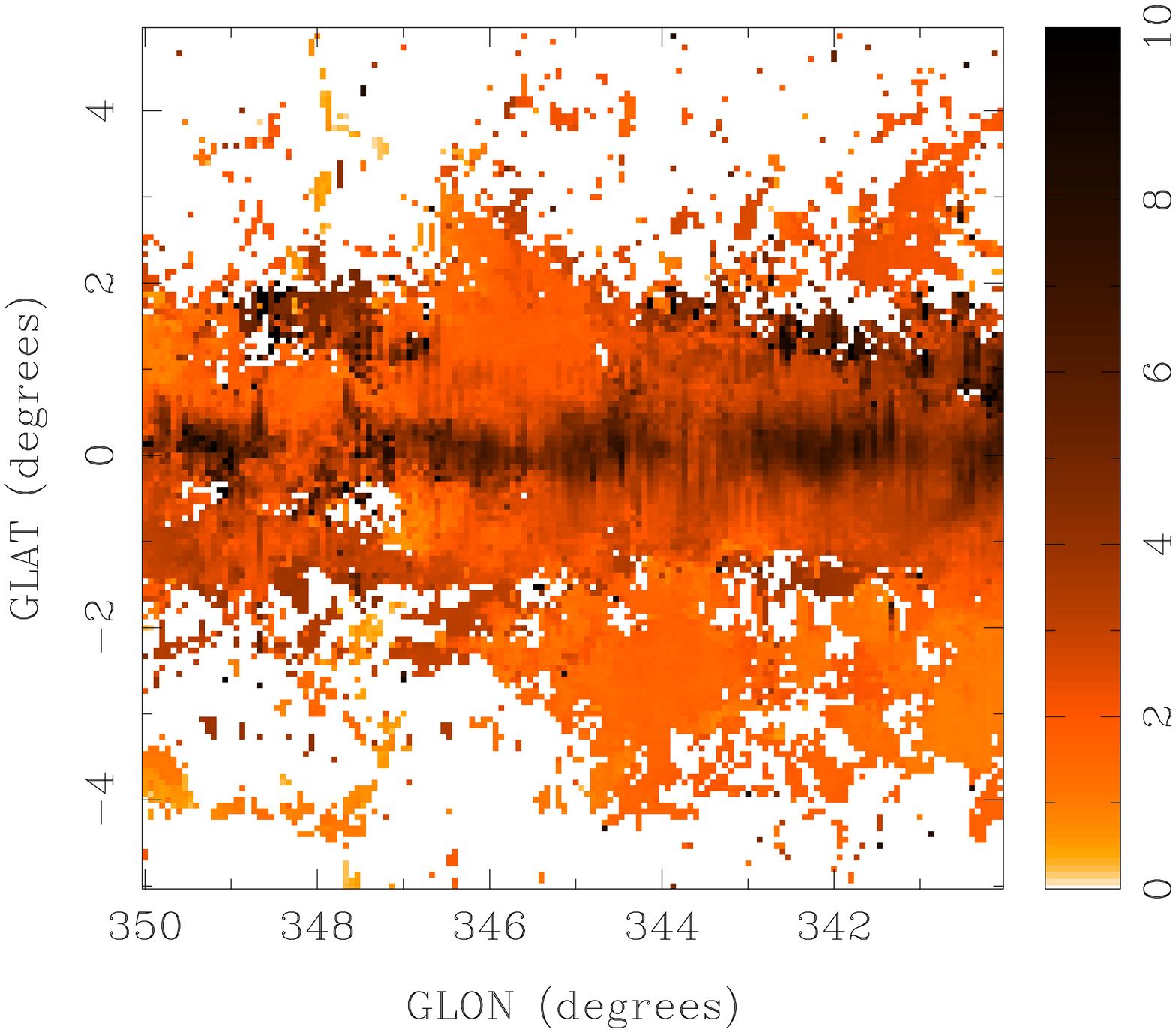}
\end{center}
  \caption{({\it Left}): Integrated intensity image for the CO (J=1--0) line emission obtained by the NANTEN telescope 
for the region $340^\circ<l<350^\circ$ and $-5^\circ<b<5^\circ$.  The intensity scale on the right is in units of 
km~s$^{-1}$~cm$^{-3}$ with the $4'$ beam located in the lower left-hand corner.  ({\it Right}): Position-distance 
image (i.e., moment 1) of the region of interest.  The intensity scale on the right is in units of kpc.  
A logarithmic scaling has been applied to both images to emphasise the low-level emission.\label{fig2} }
\end{figure*}
  
The diffusion of CRs into molecular clouds depends upon the highly uncertain diffusion coefficient. 
CRs can penetrate clouds if the diffusion coefficient inside the molecular clouds is the same as the average diffusion coefficient in the 
Galaxy, as derived from spallation measurements. CRs are effectively excluded if the diffusion coefficient is suppressed \citep{Gabici1}. However, it 
has been shown that CRs at TeV energies can diffuse into even the densest parts of molecular clouds.  Whilst GeV energy CRs can diffuse 
into typical Galactic molecular clouds \citep{Gabici2009}, they have trouble penetrating the densest parts of particular molecular clouds, 
such as the GC giant molecular cloud, Sagittarius~B2 (\citep{Jones2009} and references therein).

\subsection{Calculational procedure}

Under the assumption that the CR background in the region 
of longitude range $340^\circ<l<350^\circ$ and latitude range $-5^\circ<b<5^\circ$ 
is uniform and equal to the locally observed CR flux we 
calculate the spectral features and the longitudinal and latitudinal 
profiles of the \Gray emission. We will then 
proceed to infer the missing information concerning the distance 
from the Sun to complete the 3D view of the \Gray emission.

\subsubsection{Spectral feature of the emission}

The differential photon flux 
at Earth produced by CRs interacting with the gas from a given region in the Galaxy is
\begin{equation}
\frac{dN_\gamma}{dA dE_\gamma dt d\Omega } = \int d l_d \, \int \frac{d\sigma_{p \rightarrow \gamma}}{dE_{p}} \, n(l,b,l_d) \, J(E_{p}) \,  dE_{p} 
\label{equGammaEmissivity}
\end{equation} 
where the integrals are over the heliocentric distance, $l_d$ and over 
the energy of the protons, $E_{p}$. $n(l,b,l_d)$ is the gas density 
as function of the heliocentric coordinates, latitude, $b$, 
longitude, $l$, and $l_d$, and $J(E_p)$ is the proton spectrum. In Equation 
(\ref{equGammaEmissivity}) the spectral dependence 
of the photons emitted by CR protons, expressed in terms of the differential cross section, 
$d\sigma_{p->\gamma}/dE_{p}$, is calculated using the parametrisation of \cite{Kelner}. 
A multiplication factor of 1.5, applied to the proton spectrum, 
accounts for the contribution to the emission from heavier nuclei both in CRs and in the interstellar medium \citep{Dermer,Mori}.  
From Equation (\ref{equGammaEmissivity}) for a uniform distribution of CRs the \Gray spectrum is given as
\begin{equation}
\frac{dN_\gamma}{dA dE_\gamma dt d\Omega } = \int  dE_{p}  \, \frac{d\sigma_{p \rightarrow \gamma}}{dE_{p}} \, J(E_{p}) \,  \int d l_d \, n(l,b,l_d)\,,
\label{equGammaEmissivity2}
\end{equation} 
where the integral over the line of sight, $l_d$, defines the gas column density for a given direction in latitude and longitude. 
The average atomic and molecular hydrogen column densities for the region of the Galaxy of longitude range $340^\circ<l<350^\circ$ and latitude range $-5^\circ<b<5^\circ$ are 
$3.3 \times {10}^{21}$ cm$^{-2}$ and  $1.4 \times {10}^{22}$ 
cm$^{-2}$, respectively. 
The CR spectrum is { assumed to be} equal to the CR flux measured at the top of the Earth's atmosphere. Below ${10}^6$ GeV we 
here adopt the value of 
\begin{equation}
J(E_{p}) = 1.8 {E_{p}}^{-2.7} \, {\rm {cm}^{-2}} 
{\rm s^{-1}}{\rm {sr}^{-1}}{\rm {GeV}^{-1}} \,, 
\label{equCRspectrumSteepened}
\end{equation} 
taken from the \cite{ParticleDataGroup}. 
A change in the spectral index of the measured 
CR flux is observed above ${10}^6$ GeV, which we { do not consider} 
as we are interested in the \Gray emission below 100 TeV. 

\subsubsection{Longitudinal and latitudinal profiles of the emission}

\begin{figure*}
\begin{center}
\includegraphics[width=0.47\textwidth]{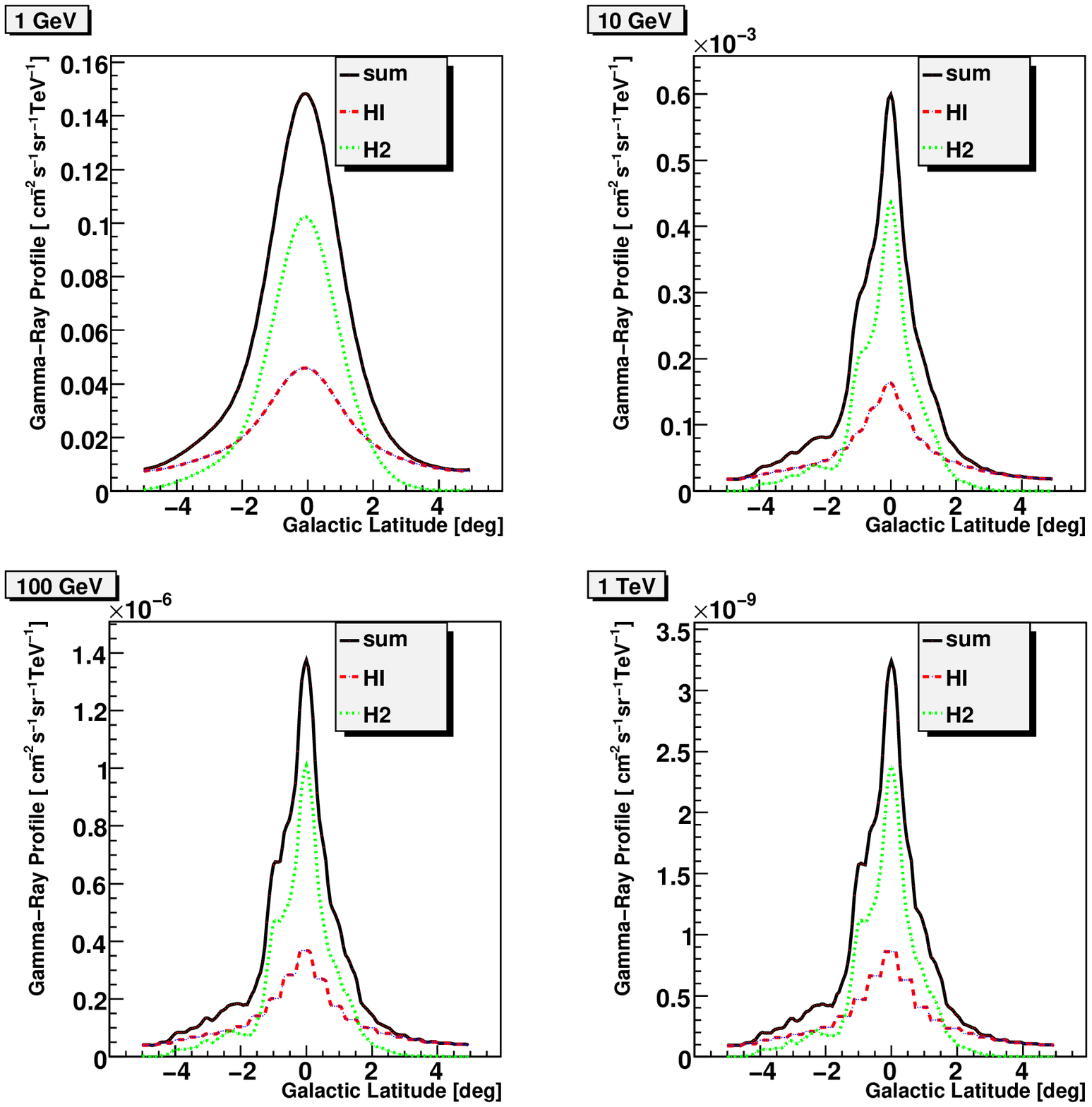}
\includegraphics[width=0.47\textwidth]{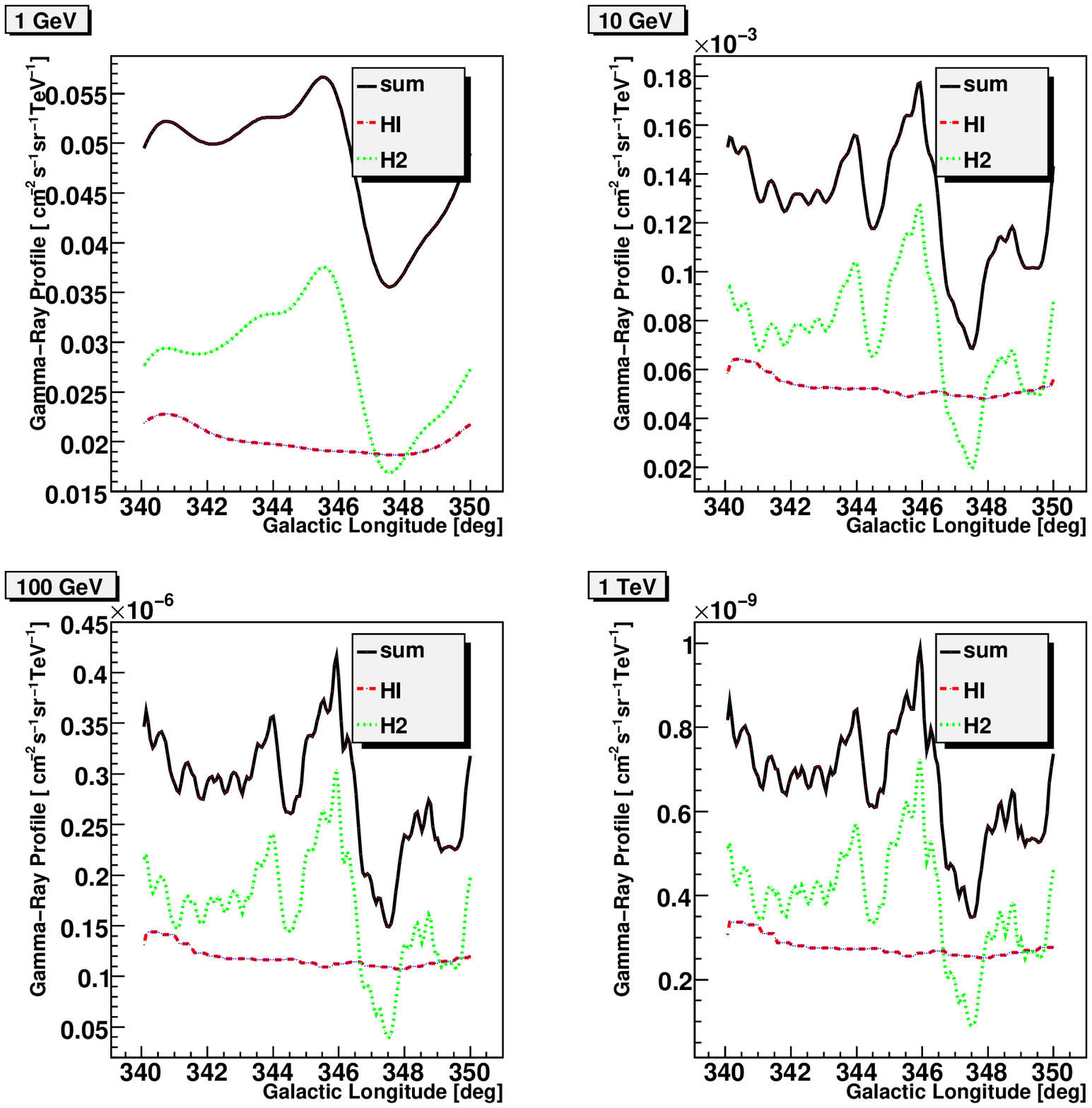}
\end{center}
\caption{ ({\it Left 4 panels}) The latitudinal profile of the 
$\gamma$-ray emission due to CRs scattering off atomic and molecular hydrogen at different energies which would be measured by Fermi and a Cherenkov telescope such as CTA 
from the region $-5^{\rm o}<b<5^{\rm o}$, integrated over the longitude range $340^{\rm o}<l<350^{\rm o}$. The emission is convolved with 
the energy dependent PSF of Fermi at 1 GeV and 10 GeV, and of a 
future Cherenkov telescope at 100 GeV and
1 TeV. The energy dependence of the Fermi PSF can be found at:{http://www-glast.slac.stanford.edu/software/IS/glastlatperformance.htm}. 
We adopt a point spread function of $0.05^\circ$ at 100~GeV and $0.02^\circ$ at 1~TeV. 
The dotted green lines are for the emission arising from CRs scattering off molecular hydrogen, the dashed red ones for the emission arising from
CRs scattering off atomic hydrogen and the solid black ones for the sum. The $\gamma$-ray profile is in units of 
cm$^{-2}$s$^{-1}$sr$^{-1}$TeV$^{-1}$. ({\it Right 4 panels}) 
The longitudinal profile of the $\gamma$-ray emission from the region $340^{\rm o}<l<350^{\rm o}$, 
integrated over the latitude range $-5^{\rm o}<b<5^{\rm o}$. 
The emission is convolved with the energy dependent PSF of Fermi at 1 GeV and 10 GeV, and of a future Cherenkov telescope such as CTA at 100 GeV and 
1 TeV. The dotted green lines are for the emission arising from CRs scattering off { molecular} hydrogen, the dashed red ones for the emission 
arising from 
CRs scattering off { atomic} hydrogen and the solid black ones for the sum. The $\gamma$-ray profile is in units of
cm$^{-2}$s$^{-1}$sr$^{-1}$TeV$^{-1}$. \label{fig4} }
\end{figure*}

The latitudinal and longitudinal profiles of the \Gray emission due to protons scattering off the atomic and molecular hydrogen from the region 
which spans Galactic longitude $340^\circ<l<350^\circ$ and Galactic latitude $-5^\circ<b<5^\circ$ are shown in Figure \ref{fig4} ({\it Left}) 
and ({\it Right}) respectively. The \Gray profiles which the Fermi telescope would observe are plotted for 1~GeV and 10~GeV. 
For a Cherenkov telescope, such as CTA, we show the \Gray profiles at 100 GeV and 1~TeV. The instrument point spread functions at 
the different energies are also taken into account. 

A peak in the emission at longitude of about $345.7^\circ$ close to the Galactic Plane is clearly visible, next to a 
dip in the longitude profile. While 
the atomic gas is generally broadly distributed along the Galactic Plane, the molecular hydrogen is less uniformly distributed and the peaks in the 
$\gamma$-ray longitude profiles correspond to the locations of highest molecular gas column density. The peaks in the 
longitudinal profile reveal the directions 
in the Galaxy where massive clouds associated with spiral arms are 
aligned along the line of sight. 
In the following discussion { we will further analyse} the region of 
latitude range $1^\circ<b<1.7^\circ$ and longitude range $345.3^\circ<l<346^\circ$, where the peak in the \Gray profiles is located.

\subsubsection{Localisation of the emission along the line of sight distance}

\begin{figure}
\begin{center}
\includegraphics[width=0.48\textwidth]{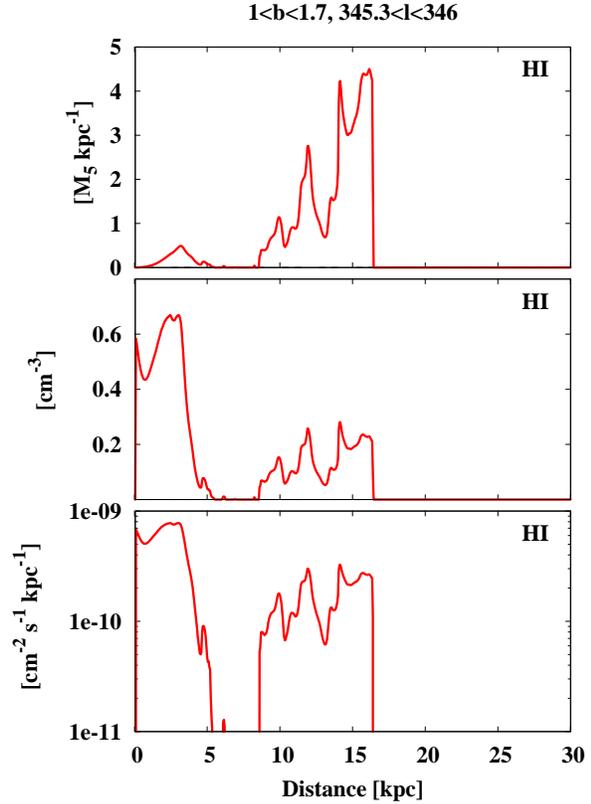}
\end{center}
  \caption{({\it Top}): Total HI mass per unit distance ($(\Delta M/\Delta d)$) in a volume element in 
$10^5$~M$_\odot$~kpc$^{-1}$ from the direction $345.7^\circ$ in Galactic longitude and $1.3^\circ$ in 
Galactic latitude is shown.  ({\it Middle}): Gas density as a function of heliocentric distance $l_d$. ({\it Bottom}): 
Unit volume emissivity above 1 GeV as a function of $l_d$ in units of 
cm$^{-2}$s$^{-1}$kpc$^{-1}$. \label{fig5}}
\end{figure}

\begin{figure}
\begin{center}
\centering
\includegraphics[width=0.48\textwidth]{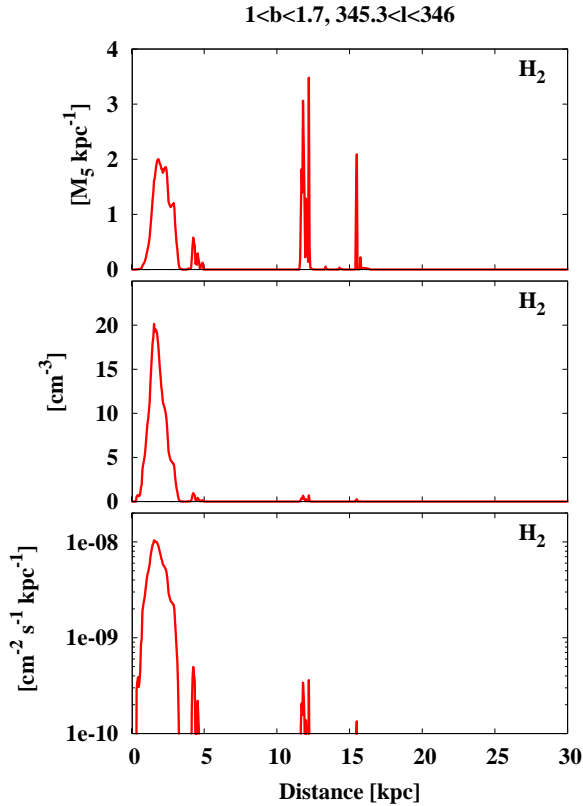}
\end{center}
  \caption{({\it Top}): Total H$_{2}$ mass per unit distance ($(\Delta M/\Delta d)$) in a volume element in $10^5$~M$_{\odot}$~kpc$^{-1}$ from the 
direction $345.7^\circ$ in Galactic longitude and $1.3^\circ$ in Galactic latitude is shown.  ({\it Middle}): Gas density as a function of distance 
$d$. ({\it Bottom}): Flux above 1 GeV in units of cm$^{-2}$s$^{-1}$kpc$^{-1}$. \label{fig6}} 
\end{figure}

In order to localise the region along the line of sight, $l_d$, 
where most $\gamma$-rays are emitted, Figures \ref{fig5} and \ref{fig6} ({\it Middle}) show 
the atomic and molecular hydrogen densities as a function of the line of sight distance, averaged over the inner  $1^\circ<b<1.7^\circ$ 
of the Galactic plane in the region of longitudes $345.3^\circ<l<346^\circ$. The value of the gas density is given every 
50~pc from 50~pc up to 30~kpc from the Sun. The density of the atomic
hydrogen as function of the heliocentric distance, $l_d$, shown in Figure 
\ref{fig5}~({\it Middle}), tends to be quite uniform. On the other
hand, the plot in Figure \ref{fig6} ({\it Middle}) shows that 
much of the molecular material is located 
within $\lesssim2-3$~kpc, which reinforces the evidence presented in Figure 
\ref{fig2}~({\it Right}). As Figure \ref{fig6}~({\it Middle}) shows, 
the molecular hydrogen density peaks to $n_{{\rm H}_2}=20$~cm$^{-3}$ in this region.

The total mass (H$_{2}$ and HI) in the element of volume, 
$\Delta V =  l_d^2  \,  cos(b)
\, \Delta b  \, \Delta l  \, \Delta l_d$, with $\Delta b=\Delta
l=0.01$ rad $(=0.7^\circ)$ and $\Delta l_d=50$ pc around the directions of
longitude $345.7^\circ$ and latitude $1.3^\circ$, is 
\begin{equation}
\label{eqnMass}
{\Delta M} = n(l,b,l_d) \, \Delta V \,.
\end{equation}
${\Delta M}$ is calculated every 50 parsecs along 
the line of the sight distance,
$l_d$, corresponding to each value of the gas density $n(l,b,l_d)$.  
In Figures \ref{fig5} and 
\ref{fig6}~({\it Top}) we plot the mass per unit distance ${\Delta M}/{\Delta l_d}$ 
in a volume element in units of M$_5$~kpc$^{-1}$ where $M_5=10^5$~M$_\odot$. 

Assuming the local CR flux, 
the unit volume emissivity above a given energy, $E^*$, is proportional 
to the total mass in the element of volume, $\Delta V$, and 
in photons~cm$^{-2}$~s$^{-1}$~kpc$^{-1}$ it is
\begin{eqnarray*}
\Delta F(E>E^*) &=& 1.2\times {10}^{-13}{\left(\frac{E^*}{1\textrm{~TeV}}\right)}^{-1.7} \\
&\times & {\left(\frac{l_d}{1 \textrm{~kpc}}\right)}^{-2} \frac{\Delta M}{M_5}\,.
\end{eqnarray*}
\begin{equation}
\label{eqn1}
\end{equation}
The emissivity per unit volume as a function of the line of sight distance, 
${\Delta F(E>E^*)}/{\Delta l_d}$, is plotted in Figures \ref{fig5} and 
\ref{fig6}~({\it Bottom}). Figures \ref{fig5} and \ref{fig6}~({\it Bottom}) show 
that most of the $\gamma$-ray
emission from the region of 
longitude $345.3^\circ<l<346^\circ$ and latitude $1^\circ<b<1.7^\circ$ is produced between 0.5 and 3 kpc. 

The combined longitudinal, latitudinal and distance profiles 
provide together a tomographical view of the $\gamma$-ray emission. 
The investigation 
of the distribution of the \Gray emission gives the locations in the Galaxy where most \Gray emission is 
produced within a reasonably 
limited region, where a single or at most few massive clouds are located, making it possible to constrain the CR flux in this particular region. 
 

\subsection{Leptonic emission and secondary electrons}
\subsubsection{Leptonic Emission: The Inverse Compton and Bremsstrahlung Contribution}
 
Inverse Compton scattering of cosmic ray electrons results in a 
non-negligible contribution to the overall diffuse $\gamma$-ray emission at 
GeV, and perhaps also at TeV energies (see e.g. \citep{Aharonian2000b}). 
However the relative contribution of this component (compared to 
the bremsstrahlung and $\pi^0$-decay $\gamma$-rays) from specific dense 
regions is significantly reduced because of the enhanced gas density. 
The ratio of the bremsstrahlung-to $\pi^0$-decay $\gamma$ 
rays does not depend on the density. However, the contribution of the 
bremsstrahlung to the total diffuse emission above 1 GeV is rather low, so 
it can be ignored. This is true also for the bremsstrahlung from secondary 
electrons. Therefore in this paper we will focus only on the $\pi^0$-decay 
component. 

\begin{figure}
\begin{center}
\centering
\includegraphics[width=0.48\textwidth]{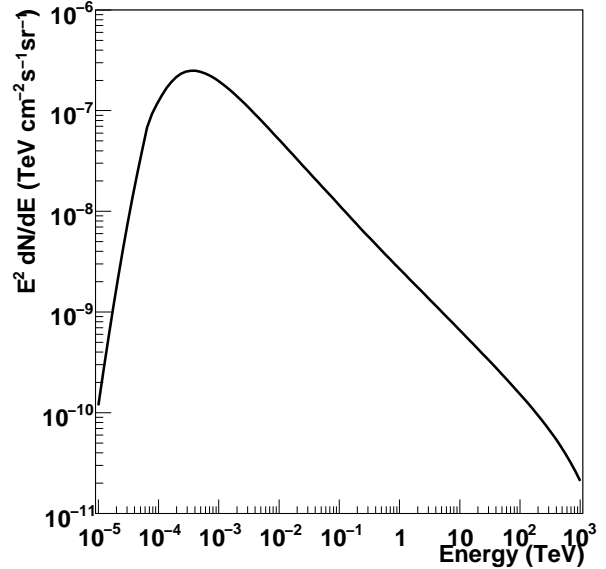}
\end{center}
\caption{The spectra of the $\gamma$-ray emission from the region $345.3^{\rm o}<l<346^{\rm o}$ and  $1^{\rm o}<b<1.7^{\rm o}$ arising from the 
observed CR spectrum as in Equation (\ref{equCRspectrumSteepened}). The atomic hydrogen column density is $2.1 \times 10^{22}$~cm$^{-2}$ and $8.0 \times 10^{22}$~cm$^{-2}$ for the molecular hydrogen.\label{fig8}} 
\end{figure}

\subsubsection{Possible Role of Secondary Electrons and Positrons}
When CRs interact with ambient matter through inelastic collisions 
they produce not only neutral pions, but also charged pions. These charged pions emit 
secondary electrons, positrons and neutrinos. 
Thus, in addition to the \Gray emissivity due to neutral pion decay, there will be a simultaneous radio synchrotron component due to the secondary 
electrons and positrons produced concomitantly with the neutral pions.  Whilst the emissivity of such electrons is beyond the scope of this paper, 
we pause only to note that their emissivity is proportional to the mass of ambient matter and the CR flux in most regions.  In the densest regions 
(i.e. $>10^4$~cm$^{-3}$) however, the synchrotron emissivity will scale as the volume because of the increased efficiency of bremsstrahlung cooling 
\citep{Jones2009}.  Since synchrotron emission from secondaries 
is not readily observable in the cases so far attempted \citep{Jones2008, Protheroe2008}, this method of checking the validity 
of the CR-{\it sea} assumptions at GeV energies is probably not feasible.

\section{Results and Discussion}\label{sec:results}

\begin{figure}
\begin{center}
\includegraphics[width=0.48\textwidth]{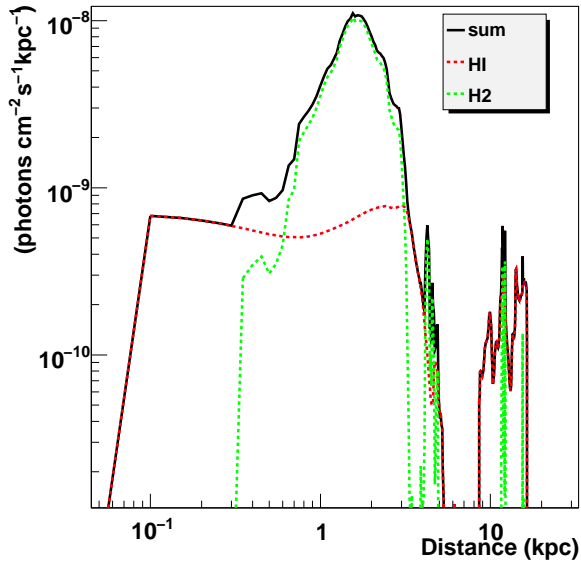}
\end{center}
\caption{Flux above 1 GeV as function of the line of sight distance 
in units of cm$^{-2}$ s$^{-1}$ {kpc}$^{-1}$, 
produced by CRs interacting with the atomic and molecular gas. The 
atomic hydrogen is uniformly distributed in the Galactic Plane, whereas 
the molecular hydrogen is more localised.\label{fig9}}
\end{figure}

Table 1 shows the total hydrogen mass from the LAB and NANTEN
surveys and the integral \Gray flux above 1~GeV and above 100~GeV as a function
of the line of sight, $l_d$,  
from the region $345^\circ.3<l<346^\circ$ and $1^\circ<b<1^\circ.7$
centred on $l=345^\circ.7$ and $b =1^\circ.3$. The values for the masses
and fluxes are obtained by integrating Equation (\ref{eqnMass}) and
(\ref{eqn1}), respectively, along given portions of the distance, $l_d$. 
It is instructive to note that 
2$\%$ of the emission 
comes from the region $d< 0.5  \, {\rm kpc}$, 85$\%$ from 
the region $0.5  \, {\rm kpc} < d< 3  \, {\rm kpc}$, and 13$\%$ from the region $d> 3  \, {\rm kpc}$. 
The peak in the $\gamma$-ray emission in Figure \ref{fig4} from the direction ${345.7}^{\rm o}$ close to 
the Galactic plane is mostly produced within 0.5 kpc and 3 kpc distance from the Sun. 
Thus the $\gamma$-ray emission from this direction provides a unique probe of the CR spectrum in 0.5-3 kpc, provided the 
$\gamma$ emission morphology is shown to match the gas in 
this distance range. In Figure \ref{fig8} we show the predicted $\gamma$-ray spectra for the same region,  
$345.3^\circ<l<346^\circ$ and $1^\circ<b<1.7^\circ$, arising from the CR spectrum in Equation (\ref{equCRspectrumSteepened}). 
As shown in Figure \ref{fig9}, 
the contribution to the emission from the atomic hydrogen in the 
$0.5  \, {\rm kpc} < d< 3  \, {\rm kpc}$ amounts to about 10$\%$ of the total contribution from atomic and molecular hydrogen. 
The emission beyond 3 kpc arises predominantly (80$\%$) from the more uniformly distributed HI gas. 

The Fermi Collaboration quotes an integral sensitivity of $5 \times {10}^{-10}$ photons~cm$^{-2}$~s$^{-1}$ above 1 GeV from a 
region which has an angular resolution of 0.7 by 0.7 degrees. This makes the detection of the predicted emission for Fermi possible. Future Cherenkov 
telescopes, such as CTA, will need to reach a sensitivity of about 5 $\times {10}^{-12}$ photons~cm$^{-2}$~s$^{-1}$ 
for observing a region of 0.7 by 0.7 degrees above 100 GeV. 

\begin{table*}
\begin{minipage}{160mm}
\begin{center}
\caption{Summary of parameters as a function of distance for the region 
$345.3^\circ<l<346^\circ$ and $1^\circ<b<1.7^\circ$}
 \begin{tabular}{llllll}
  \hline
   Region &  Mass & Mass (H$_2$) &  Mass (HI) & F( E $>$ 1 GeV)   
&  F(E $>$ 100 GeV)   \\
  (kpc) & {(M$_{5}$  )}  &  (M$_{5}$) &  (M$_{5}$) & (cm$^{-2}$~s$^{-1}$) 
& (cm$^{-2}$~s$^{-1}$) \\ 
   (1) & (2) & (3)& (4) & (5) & (6)\\
\hline
    0.5 $<$ d $<$ 30 & 20.2   & 4.2   & 16.0         
& 1.7 $\times {10}^{-8}$  &  6.8 $\times {10}^{-12}$ \\
    d $<$ 0.5        & 0.0015 & 0.0005 & 0.001       
& 2.9 $\times {10}^{-10}$ &  1.2 $\times {10}^{-13}$  \\
    0.5 $<$ d $<$ 3  & 3.2    & 2.8   &  0.4         
& 1.4 $\times {10}^{-8}$  &  5.7 $\times {10}^{-12}$  \\
    d $>$ 3          & 17.0   & 1.4   & 15.6         
& 2.3 $\times {10}^{-9}$  &  9.1 $\times {10}^{-13}$ \\
\hline
\end{tabular}
\end{center}
\end{minipage}
\label{table:params}
\end{table*}

We have here presented a methodology to test the cosmic ray flux 
in discrete distant regions of the Galaxy by combining the most recent 
high resolution \Gray and gas data. By assuming 
a flat rotation curve model 
of the Galaxy we have estimated the gas number density at discrete positions 
as a function of the heliocentric latitude, longitude and distance.
Through the knowledge of the gas number density 
we have determined the total mass of the gas in discrete regions of the Galaxy 
and there we have predicted the \Gray emissivity under the assumption that the CR flux 
is equal to that measured at Earth. By comparing the predicted and the 
measured \Gray flux we have obtained a test of 
the level of the CR flux in these regions. The CR flux has been calculated as a function of the Galactic longitude, latitude and 
heliocentric distance, providing a sort of {\it tomography} of the CR flux in different locations of the Galaxy.

For our investigations we have used the higher resolution $^{12}$CO (viz. $4'$ NANTEN  data) 
and LAB HI surveys.  The resolution 
of the LAB HI data used in the present investigation is 0.6 deg. 
The high resolution of the 
NANTEN data allows for the first 
time detailed comparisons of our $\gamma$-ray 
predictions with new TeV and multi-GeV data. In future we will make use of the higher resolution (2 arcmin) 
Southern Galactic Plane HI Survey 
available at \footnote{\url{http://www.atnf.csiro.au/research/HI/sgps/queryForm.html}}, in 
order for our predictions 
to match with higher resolution TeV data in cases where the 
$\gamma$-ray component from HI is similar 
to or dominates over the $\gamma$-ray component from  H$_{2}$ 
(e.g from Figure \ref{fig5} in the 10 to 15 kpc range).

\subsection{Uncertainties in the CR flux estimate}

Uncertainties in the distance estimates affect the calculation of the CR flux in Equation (\ref{eqn:fluxCloud}) in three ways. 
Firstly, in the determination of the conversion factor X in Equation 
(\ref{eq:X}), 
by which the emissivity of the CO line is converted to ambient matter density. 
As previously discussed, the uncertainty in the $X$ factor is 50$\%$ \citep{arimoto}. 
Secondly, there are uncertainties in the determination of the number  
density of H$_2$ molecules from the measured CO intensity.   
Specifically, the error on the distance (especially at small 
distances;  c.f. Fig.\ref{fig1}) is as large as  
2~kpc.  We note however, that for extended sources (i.e., sources  
whose intrinsic size is larger than the beam size), Equation 
(\ref{eqn:fluxCloud}) shows that the CR flux density is proportional to  
the distance squared and inversely proportional to the mass.  Since  
the mass scales as the distance square, the relative errors exactly  
cancel.  For unresolved sources (i.e., sources whose intrinsic size is  
smaller than the beam size), we note that the above errors still hold,  
however, there will be a systematic underestimation of the true CR  
flux density because of beam dilution.  Thus estimates for the CR flux  
density need to be estimated using {\it extended} sources only.

It is instructive to note that only 85$\%$ of the emission, which would be measured 
from the direction $345.3^\circ<l<346^\circ$ and $1^\circ<b<1.7^\circ$, 
is produced in the region $0.5  \, {\rm kpc} < d< 3  \, {\rm kpc}$. Therefore 
only this 85 $\%$ of the emission is to enter in Equation 
(\ref{eqn:fluxCloud}) in order to 
estimate the CR flux in the region $0.5  \, {\rm kpc} < d< 3  \, {\rm kpc}$. 
Moreover, part of the \Gray emission, which would be measured from the 
region $345.3^\circ<l<346^\circ$ and $1^\circ<b<1.7^\circ$, is produced 
through competing IC and non-thermal bremsstrahlung mechanisms of primary and secondary electrons. In particular, in the energy range between 10 MeV and 100 MeV 
bremsstrahlung processes of background and secondary electrons produce 
a non negligible contribution to the \Gray emission, so that the 
presented method can be applied to the analysis of the \Gray emission above 100 MeV-1 GeV, where moreover 
the angular resolution of the Fermi LAT detector 
is substantially better. Finally unknown CR source(s) might produce an enhancement of the local CR background. 

\section{Conclusions}\label{sec:conclusions}
 
{ The CR level in distant regions of the Galaxy 
and the distribution of Galactic CR sources are unknown.
\Grays emitted when CR protons and nuclei interact with the atomic and molecular 
gas have been long recognised as a unique probe of the CR flux and spectrum in 
the Galaxy \citep{Ginzburg,Puget,Casse,BloemenB,BloemenC,Lebrun,Bertsch,Strong}.

We have here introduced a methodology which aims to provide a test bed 
for current and future \Gray observatories to 
explore the cosmic ray flux at various positions in our Galaxy. 
In order to investigate the CR spectrum in distant regions of the Galaxy we have here assumed 
that the CR flux and spectrum measured at the Earth is representative 
of the typical CR flux and spectrum present throughout the Galaxy. The \Gray emission 
arising from such CR spectrum has been calculated in specific regions of the Galaxy, using the 
knowledge on the distribution of the atomic and molecular gas, as given by the LAB and Nanten surveys, respectively. 
%This is equivalent to assume that 
%firstly it exists a definable CR {\it sea} of homogenised and isotropised CRs, 
%far from any of the uniformly and continuously distributed CR sources, 
%which is the typical CR flux and spectrum in the Galaxy, and secondly that the Earth 
%sits in a typical region far from strong sources and where this {\it sea} 
%can be measured directly.  Due to the diffusion times of the CRs within the Galaxy, 
%there seems to be little doubt that a CR {\it sea} exists. However, 
%CR gradients with respect to the {\it sea} level are expected 
%on scales of hundreds of parsecs close to CR sources, and possibly also 
%on 1-2~kpc scales, at low energy (below GeV) if the distribution of sources 
%is not homogeneous and the CR injection spectra vary in different sources. 
%In this regard it is questionable that the Earth 
%sits in a typical region of the Galaxy far from CR sources, 
%where the {\it sea} can be measured directly. 
The peaks in the longitudinal 
and latitudinal distributions of the gas correspond to directions in the Galaxy, 
where only one or at most few massive clouds contribute to the total mass. This can be 
demonstrated by studying how the gas is distributed along the line of sight distance. 
We have then exploited this knowledge of the gas number density point 
by point to determine the total mass of the gas of some MCs 
of the Galaxy. From the total mass 
of the atomic and molecular gas we have calculated the level of the \Gray emission in such gas clouds. 
The predicted \Gray emission from such gas 
clouds can be compared with observations of present and future $\gamma$-ray detectors 
and provide a test of the CR density in distant locations of the Galaxy. 
We have, in particular, demonstrated that the new high quality data of NANTEN and the performance of the Fermi LAT telescope are
sufficient for meaningful probes of CR densities in specific 
localised regions.}

Equation (\ref{eqn:fluxCloud}) shows that the CR flux density is proportional to  
the distance squared and to the \Gray ray flux and inversely proportional 
to the mass of the MC. The error in the determination of distances (especially at small 
distances; c.f. Fig.\ref{fig1}) is as large as 2~kpc.  We note however, 
that for extended sources (i.e., sources  whose intrinsic size is larger than the beam size) 
the mass scales as the distance square and the relative errors on mass and distance exactly  
cancel. Moreover, above 100 MeV - 1 GeV, the competing leptonic processes, 
bremsstrahlung and IC of background and secondary electrons, 
produce a negligible contribution to the \Gray emission. The emission from molecular 
clouds, located far from strong CR sources, 
is then the footprint of the CR {\it sea} within the cloud and molecular clouds serve then 
as {\it barometers} of the CR pressure. The detection of 
even a single under-luminous 
cloud with respect to the predictions based on the local CR density 
would cast severe doubt on the assumption that the level of the observed 
CR flux is representative of the average CR flux and spectrum 
in the Galaxy. This conclusion is even reinforced if additional gas, not yet revealed, 
is present in the cloud. In this regard, by comparing various interstellar gas 
tracers, EGRET $\gamma$-ray sky maps, HI and CO lines but also dust thermal emission, 
\cite{Grenier} unveiled clouds of dark gas surrounding the nearby CO clouds. \cite{Grenier} concluded that the 
additional mass from dark clouds implies that the cosmic ray densities are overestimated in 
regions rich in dark clouds. 

A detailed analysis of the spectral features of the \Gray emission from various regions 
of the Galaxy is also crucial. In fact, deviations from the standard spectrum 
are indications of local CR sources. The regions in the Galaxy where the spectral features 
and/or the normalisation of the $\gamma$-ray spectrum 
cannot be explained by interactions of a CR spectrum similar 
to that measured close to the Sun, are of interest to 
probe the sites of Galactic CR injection, such as SNRs and pulsars. 

The Fermi observatory, which detects \Grays in the energy range between tens of MeVs and hundreds of GeVs and which 
benefits by a large effective area and good angular resolution, is an ideal instrument to perform survey of 
extended diffuse sources such as molecular clouds in the Galaxy. Additionally, future VHE detectors, 
such as CTA, will offer improved sensitivity and angular resolution, 
close to the angular resolution of the NANTEN survey, and could investigate the morphology of the \Gray 
emission and compare it to the structures of molecular hydrogen clouds in the Galaxy. Future IACT telescopes 
ability to resolve clouds will of course strongly depend upon the instrument sensitivity to detect extended sources. 
On the other hand, future large field of view experiments, such as HAWC, will be able to investigate 
the level of the cosmic ray background on large portions of the sky \citep{Sinnis}.

Finally, we note that we have applied our analysis of the spectral line observations to 
predict the $\gamma$-ray emission in only one distant region. It
serves mostly as a 'proof-of-concept', to be followed by a more
comprehensive study of the entire Galaxy.

\section*{Acknowledgments}

{ First of all we would like to thank the anonymous referee for
his/her help to improve the manuscript. }

The NANTEN telescope was operated based on a mutual agreement between Nagoya University and the Carnegie Institution of Washington. We also  acknowledge that the operation of NANTEN was realised by contributions from many Japanese public donators and companies. This work is financially supported in part by a Grant-in-Aid for Scientific Research from the Ministry of Education, Culture, Sports, Science and  Technology of Japan (Nos. 15071203 and 18026004, and core-to-core  program No. 17004) and from JSPS (Nos. 14102003, 20244014, and  18684003). 

S. Casanova and S. Gabici acknowledge the support from the European Union under Marie Curie Intra-European fellowships. We want to thank Dr Kester Smith and Joanne Dawson for very helpful discussions.

\end{document}